\documentclass[conference]{IEEEtran}
\IEEEoverridecommandlockouts
\usepackage{cite}
\usepackage{amsmath,amssymb,amsfonts}
\usepackage{algorithm}
\usepackage{algpseudocode}
\usepackage{graphicx}
\usepackage{textcomp}
\usepackage{xcolor}
\usepackage{listings}
\def\BibTeX{{\rm B\kern-.05em{\sc i\kern-.025em b}\kern-.08em
    T\kern-.1667em\lower.7ex\hbox{E}\kern-.125emX}}
\begin{document}

    \title{Flow Optimization at Inter-Datacenter Networks for Application Run-time Acceleration}

\author{
    \IEEEauthorblockN{Berta Serracanta}
    \IEEEauthorblockA{
    \textit{UPC BarcelonaTech}\\
    Barcelona, Spain \\
    berta.serracanta@upc.edu}
    
    \and
    \IEEEauthorblockN{Alberto Rodriguez-Natal}
    \IEEEauthorblockA{
    \textit{Cisco Systems}\\
    San Jose, USA \\
    natal@cisco.com}
    
    \and
    \IEEEauthorblockN{Fabio Maino}
    \IEEEauthorblockA{
    \textit{Cisco Systems}\\
    San Jose, USA \\
    fmaino@cisco.com}
    
    \and
    \IEEEauthorblockN{Albert Cabellos-Aparicio}
    \IEEEauthorblockA{
    \textit{UPC BarcelonaTech}\\
    Barcelona, Spain \\
    alberto.cabellos@upc.edu}
}

\maketitle

\begin{abstract}

In the present-day, distributed applications are commonly spread across multiple datacenters, reaching out to edge and fog computing locations. The transition away from single datacenter hosting is driven by capacity constraints in datacenters and the adoption of hybrid deployment strategies, combining on-premise and public cloud facilities. However, the performance of such applications is often limited by extended Flow Completion Times (FCT) for short flows due to queuing behind bursts of packets from concurrent long flows. To address this challenge, we propose a solution to prioritize short flows over long flows in the Software-Defined Wide-Area Network (SD-WAN) interconnecting the distributed computing platforms. Our solution utilizes eBPF to segregate short and long flows, transmitting them over separate tunnels with the same properties. By effectively mitigating queuing delays, we consistently achieve a 1.5 times reduction in FCT for short flows, resulting in improved application response times. The proposed solution works with encrypted traffic and is application-agnostic, making it deployable in diverse distributed environments without modifying the applications themselves. Our testbed evaluation demonstrates the effectiveness of our approach in accelerating the run-time of distributed applications, providing valuable insights for optimizing multi-datacenter and edge deployments.
\end{abstract}

\begin{IEEEkeywords}
distributed applications, flow optimization, SD-WAN
\end{IEEEkeywords}

\section{Introduction}

In today's datacenter landscape, applications have undergone a significant transformation. They have transitioned from hosting applications within a single datacenter to adopting a distributed model that spans multiple datacenters and extends to the edge and the fog. This change is driven by different reasons, first datacenters are operating at capacity and the natural shift is pushing workloads to the edge. And second, more and more companies are adopting a hybrid deployment strategy, distributing workloads on-premise and public cloud facilities. This trend is happening due to a number of reasons such as security protection for critical workloads, increased flexibility and regulatory and data sovereignty requirements. In this context, emerging applications are running on a strongly distributed computing platform: fog, edge and multiple datacenters including public cloud and on-premise datacenters. 




During the past decade, numerous efforts have been dedicated to optimizing the network \emph{within a datacenter}, resulting in a wide range of novel techniques. Notable examples are pFabric, which introduces a scheduling algorithm to prioritize short flows for improved of latency-sensitive applications, and the "Data center TCP (DCTCP)" \cite{DCTCP}, which introduces a congestion control mechanism addressing the needs of datacenter environments.


One relevant insight of such extensive research is that the round-trip time (RTT) of short flows correlates with the application run-time. This means that in distributed applications short flows are used to communicate its different components, and the performance of such flows is critical towards the overall performance of the application. A notable example of this are Remote Procedure Call (RPC), short flows (typically below 256kb) that are extensively used in distributed applications. As a result of this insight, there are many proposals in datacenter networks to prioritise short flows in front of long flows.

It has been reported that short flows have the potential to complete within 10-20 microseconds \cite{pfabric}, but they often experience delays of tens of milliseconds. This delay is primarily caused by the queuing of short flows behind bursts of packets generated by large flows from concurrent workloads such as backup, replication, and data mining. These bursts significantly increase the completion times of short flows, leading to substantial latency discrepancies from their expected durations and larger application run-times.



When considering the current environment where distributed applications run on top of multiple datacenters (public/on-premise) and at the edge, the network interconnecting such computing platforms becomes the bottleneck for short flows, this network is typically referred as the Software-Defined Wide-Area Network (SD-WAN). For instance and as reported by Google, in environments with mixed traffic, WAN traffic reacts to congestion 10-10,000 times later than datacenter traffic due to the difference in RTTs, becoming the bottleneck \cite{annulus}.



In this paper we aim to accelerate the run-time of distributed applications deployed on multi-datacenter and edge environments interconnected via an SD-WAN fabric. To achieve this we prioritize at the WAN level short flows over long flows. Our proposed solution works with encrypted traffic, which is a strong requirement towards deployability in today's Internet. To assess its effectiveness, we evaluate our solution on a testbed comprising 4 routers that connect a remote office suite \cite{nextcloud} to its clients using two distinct tunnels. By effectively segregating long and short flows, we consistently achieved a 1.5 reduction times Flow Completion Time for short flows.

\section{Proposed Solution}

The solution prioritises short over long flows and aims to minimize the Flow Completion Time (FCT) and tail latency for short flows in multi-datacenter deployments. The ultimate goal is to reduce application response time since it has been reported that response time and FCT correlate \cite{response_FCT_relation}. 

\subsection{Network Scenario}

\begin{figure}
        \includegraphics[width=\columnwidth]{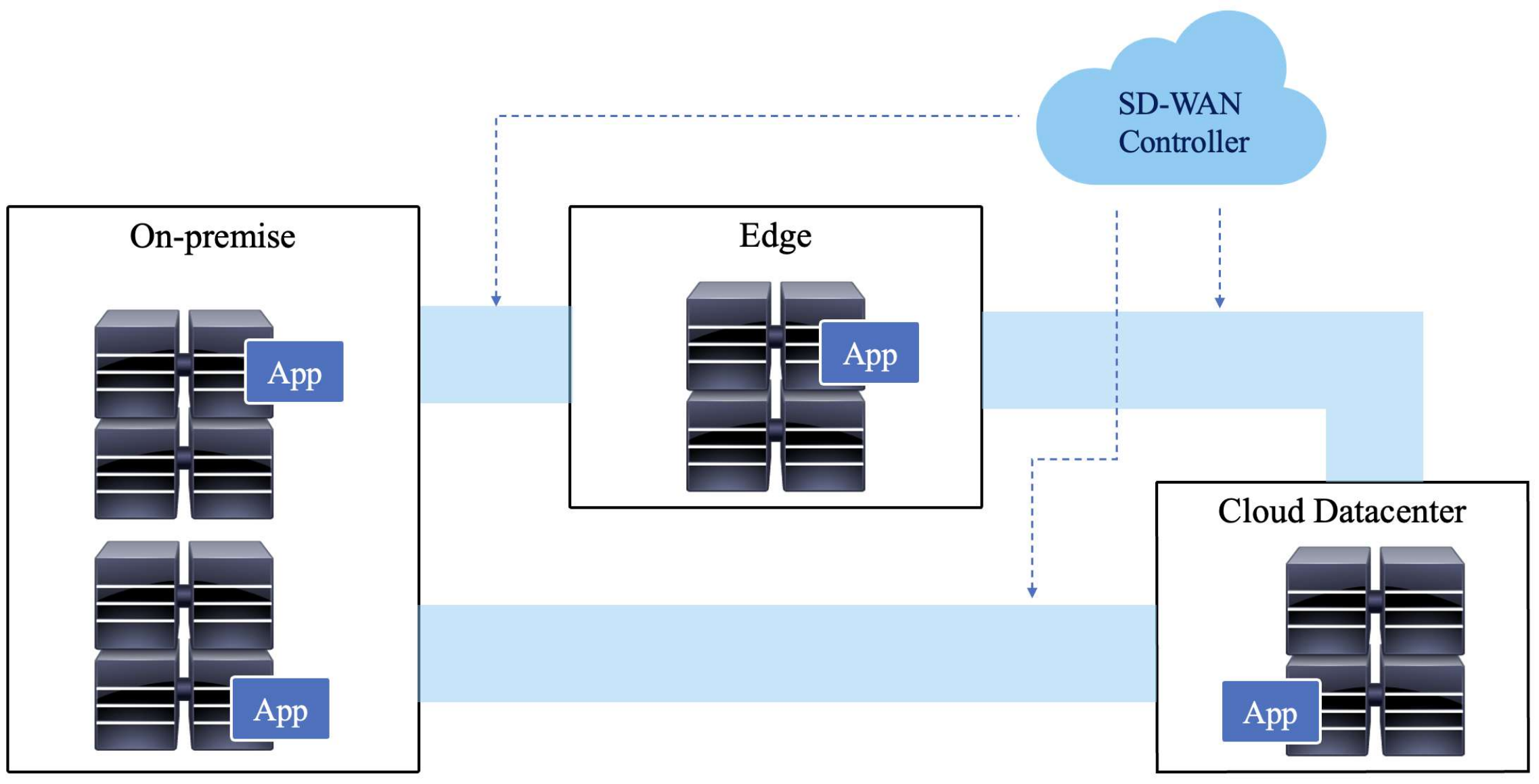}
        \caption{Scenario}
        \label{gen_scenario}
    \end{figure}

Our scenario (Fig.~\ref{gen_scenario}) is centered around a highly distributed application. This application is composed of multiple microservices, each capable of being deployed in various locations, including on-premise, edge, and public/private cloud facilities. The connectivity between these locations is established through a Software-Defined WAN (SD-WAN).

\vspace{1mm}

\subsection*{Proposal}
We propose separating long and short flows traversing Wide-Area Networks (WANs), this separation is achieved using a splitter that distinguishes packets as part of a long or short flow depending on the total number of packets composing the flow. Once the packets are marked accordingly, we leverage the potential of SD-WAN to route them through two different tunnels with the same properties. This allows short flows, which are typically more critical, to avoid queue waiting due to packet bursts from longer flows, reducing their FCT.

A very important design consideration is that this solution needs to work with encrypted traffic, since this is common-practice in such network scenario. This restricts our solution to work only on the available information at the IP header, since the header is sent in plain-text.

Equally important is the need for the solution to be application agnostic. This means that we do not consider extending the network-to-application interface. While many research efforts have been devoted into proposing extensions to this API (e.g., \cite{alto, qsockets}), none of these solutions have achieved wide-spread deployment. As a result, our solution is agnostic and does not require modifying the application or engaging with the developer community.

\begin{figure}
        \includegraphics[width=\columnwidth]{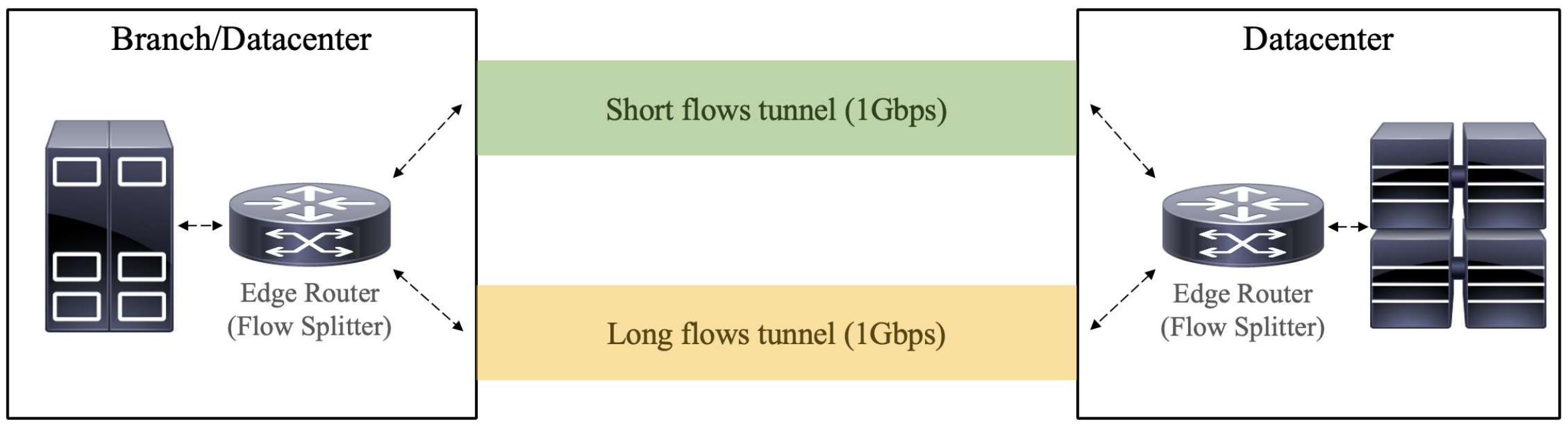}
        \caption{Overview of the implementation}
        \label{scenario}
    \end{figure}

Fig.~\ref{scenario} shows an overview of where the solution is deployed. This is a distributed application deployed across a hybrid scenario interconnected via an SD-WAN. The WAN has two distinct tunnels defined with the same properties, each one of them reserved for short and long flows, respectively.

\vspace{3mm}

\subsection{Implementation}

In order to prototype our solution we take advantage of eBPF (Extended Berkeley Packet Filter) and eXpress Data Path (XDP) \cite{ebpf}. eBPF enables line-rate programmable packet processing within the Linux kernel. This technology should allow us to split long and short flows over different WAN tunnels while introducing negligible delay.

The mechanism used to identify long and short flows works as follows. The first packet of each flow is classified as belonging to a short flow, then the WAN border router (the router interconnecting the datacenter with the client or another datacenter/edge) counts the packets being transmitted by each flow. If this count goes over a certain threshold, then the flow is classified as long. Recall that short and long flows are transmitted over different WAN tunnels (short and long WAN tunnels). As a consequence, all flows start being transmitted over the 'short tunnel' and only after they reach a certain size (in packets) they are moved to the 'long tunnel'.

\vspace{3mm}

\begin{algorithm}
\caption{Flow Splitter Algorithm}\label{alg:splitter}
\begin{algorithmic}[1]

\Require $table$  \Comment{BPF\_HASH(5tuple, pkt\_count)}
\Require $T$ \Comment{Flow size threshold}
\vspace*{1mm}
\For{Each TCP packet}
    \vspace*{1mm}
    \If{$5tuple$ not in $table$}
        \State Add $table(5tuple, 0)$
    \EndIf
    \vspace*{1mm}
    \State $table(5tuple)\rightarrow pkt\_count ++$
    \vspace*{1mm}
    \If{$table(5tuple) \rightarrow pkt\_count \geq T$}
        \State Change packet ToS
        \State Compute packet checksum
    \EndIf
    \vspace*{1mm}
    \State Return XDP\_PASS
    \vspace*{1mm}
    \EndFor
\end{algorithmic}
\end{algorithm}

In order to implement this method at the WAN border router we take advantage of eBPF \cite{ebpf}. Algorithm~\ref{alg:splitter} shows the pseudo-code of the eBPF splitter. We maintain a table that counts the number of packets of each active flow indexed by its 5-tuple. To keep the table at a reasonable size flows are removed from the table if they don't transmit a packet in a 30 seconds period. In order to choose the tunnel over which each flow is being transmitted (short or long tunnel) we mark them using the ToS field. Before leaving the eBPF splitter each packet is tagged with ToS=short or ToS=long. Then the Linux kernel is configured to forward packets according to the ToS field. 

The eBPF splitter is configured only to process TCP flows. The reason behind this is that we aim to accelerate the application run-time, and this typically depends on the speed of control flows that are short and use TCP. A notable example of this are RPCs. Other protocols remain unaffected and thus, are transmitted over the short WAN tunnel.

Finally, the resulting application run-time acceleration depends on the threshold that differentiates short from long flows. The threshold can be configured depending on the characteristics of the deployment, a more detailed discussion can be found in section ~\ref{sec:results}.

\section{Results}
\label{sec:results}
In this section, we present the evaluation of our solution, which involves the implementation of two identical tunnels that establish a connection between an on-premise datacenter and a cloud datacenter. Our network testbed, along with the properly configured eBPF splitter, forms the foundation for the four comprehensive experiments conducted to assess the performance of our proposed solution. Through these experiments, we successfully demonstrate a significant reduction in Flow Completion Time for short flows.

\subsection{Testbed}
    \begin{figure}
        \includegraphics[width=\columnwidth]{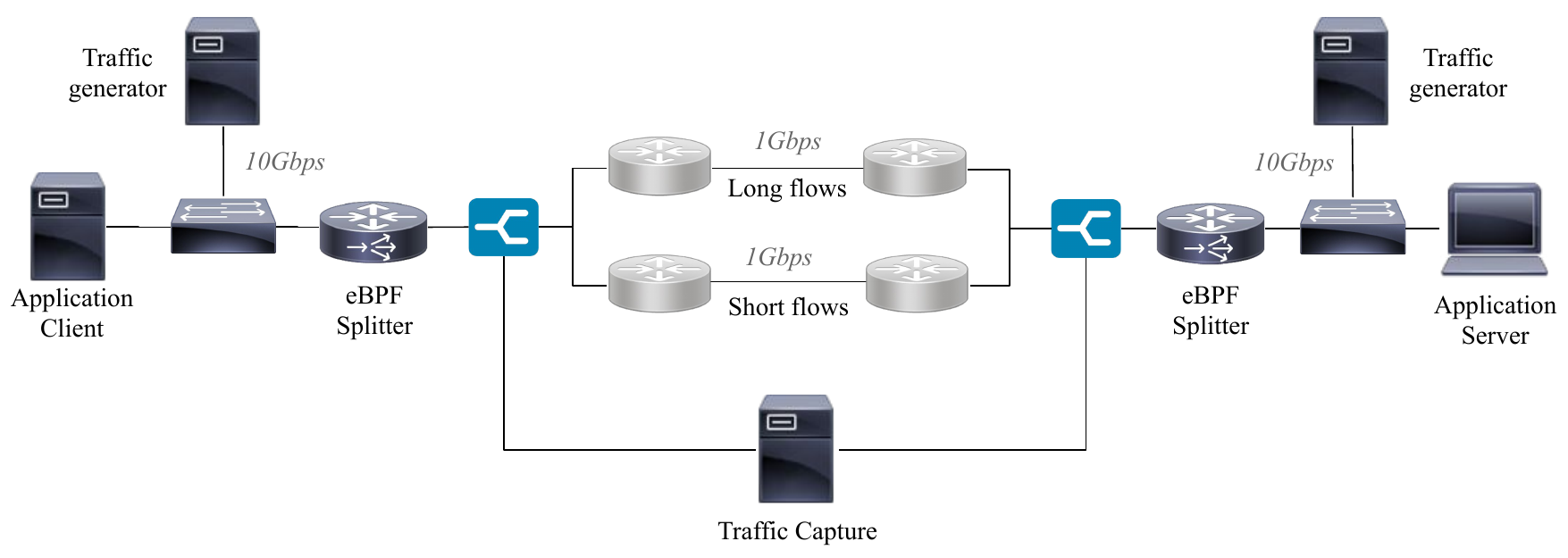}
        \caption{Testbed}
        \label{fig:testbed}
    \end{figure}
In this subsection we describe the testbed used to evaluate the performance of our solution. With this testbed in Fig.\ref{fig:testbed}, we recreated a hybrid scenario deployment interconnected via SD-WAN. We used a popular file hosting application, with the file server deployed on one side of the SD-WAN and the clients accessing it from the other end. The routers interconnecting the datacenter with the SD-WAN, and the SD-WAN with the client have configured our eBPF splitter. Our SD-WAN for the evaluation consists of four hardware routers, two for each of the tunnels defined (long and short). 

We use NextCloud \cite{nextcloud} as our test application since it is a popular application used for storing files on the cloud and for collaborative access and modifications. It allows us to perform a variety of requests, ranging from small tasks like listing folder contents ($\sim$5 packets) to larger tasks such as playing a video ($\sim$60,000 packets) or uploading big files ($>$100,000 packets). Throughout our experiment, these queries are continuously carried out following a Poisson distribution. NextCloud is deployed using the official docker image together with MariaDB \cite{mariadb} as the storage system, providing file access via API and web browser. The server we use for the application deployment is equipped with an Intel X710-DA4 that has four 10Gbps NICs and an AMD Ryzen 9 5950X CPU.

Each of the WAN border routers configured with the eBPF splitter consists of a server with the same characteristics. The SD-WAN connecting the application client and server is formed by four hardware Cisco ASR 9000v routers, each utilizing 2 ports at 1 Gbps, creating two independent and identical network paths. 

Two background traffic generators are attached to the testbed routers through WS-C3850-12X48U switches at 10Gbps ports. This additional bi-directional long flow traffic is generated using MGEN \cite{mgen}, an open-source software, to represent other WAN traffic generated from other sources. In particular, 1472B messages are generated at rates ranging between 10 to 40\% of the total capacity offered by both tunnels, following a Poisson distribution as well.

Each experiment runs for 2 minutes, with the initial 5 seconds discarded to avoid the transient state. The traffic generation is carried out on identical servers as NextCloud.

In order to measure the delay we use an optical splitter at the fibers that interconnect both SD-WAN gateways. With this we create an exact copy of the traffic both at ingress and egress. This traffic is processed by an AMD Ryzen 9 5950X CPU equipped with a set of Mellanox ConnectX-5 cards. These cards support hardware timestamping with microsecond precision, which is a necessary feature to measure delays accurately. We use DPDK to process the traffic and calculate the delay.

    \begin{figure}[btp]
        \includegraphics[width=\columnwidth]{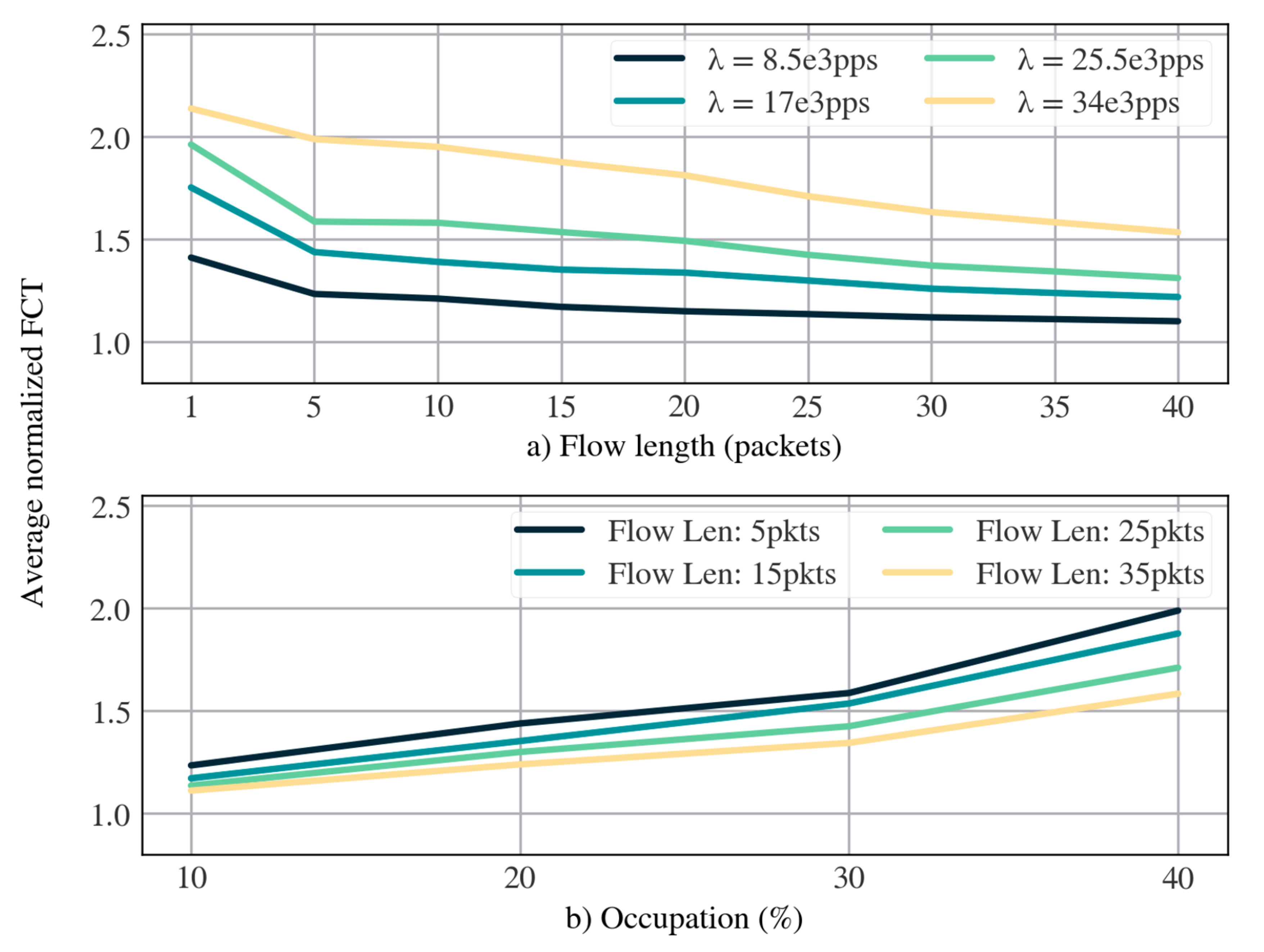}
        \caption{Effect of the eBPF splitter on accelerating Flow Completion Time.}
        \label{fig:performance}
    \end{figure}

\subsection{Performance}

    \begin{figure*}
        \includegraphics[width=\textwidth]{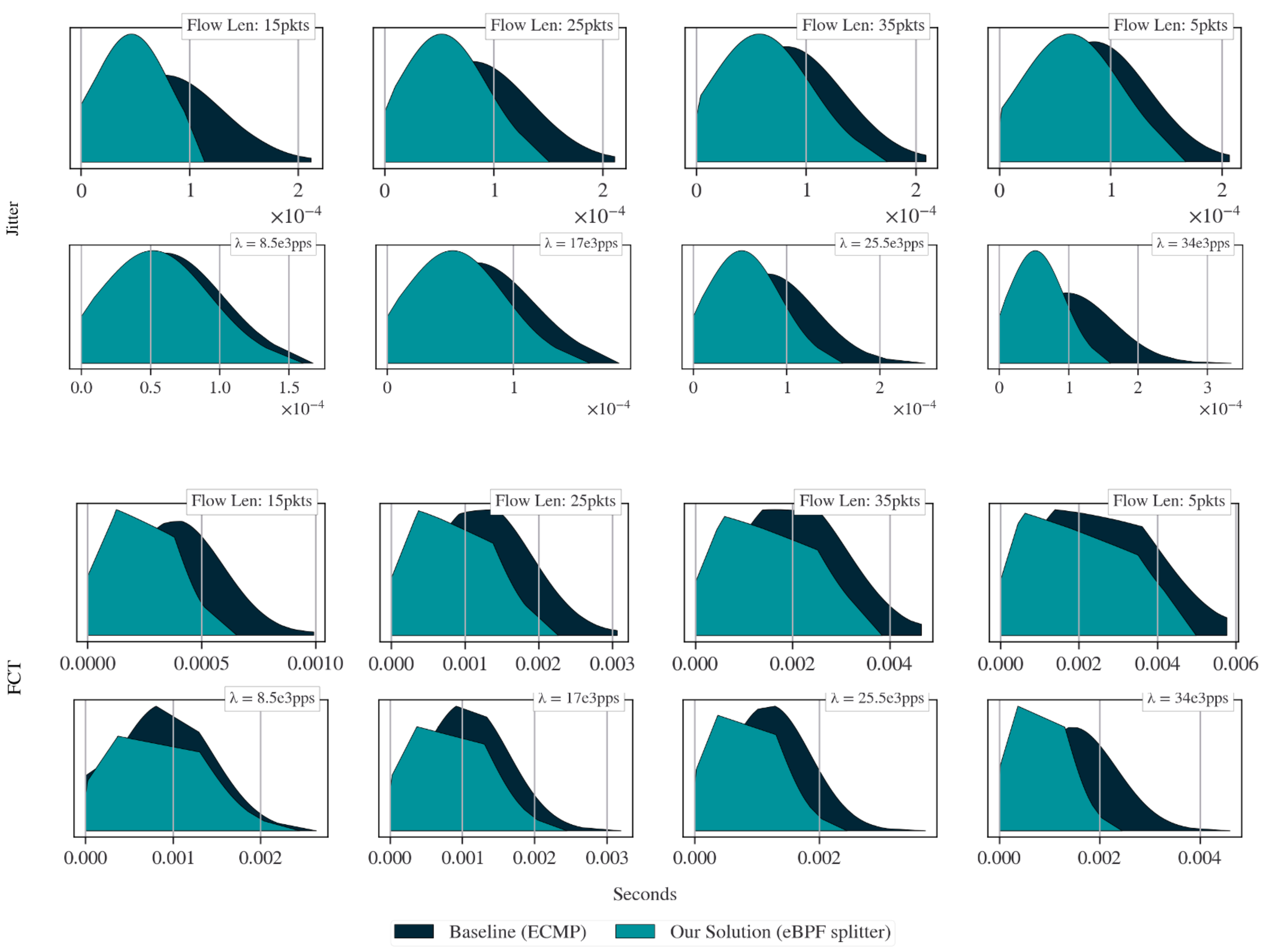}
        \caption{Probability Density Functions illustrating the effects of varying flow lengths and packet transmission rates on network jitter and Flow Completion Time.}
        \label{fig:pdf}
    \end{figure*}

In this subsection, we present the performance of our solution.

The experiments are benchmarked against Equal Cost Multi Path (ECMP) routing solution, which evenly distributes packets among the two SD-WAN tunnels. We use ECMP as baseline to compare the performance of the solution because ECMP uses the bandwidth of both tunnels and thus, all the available resources. Also note in this scenario ECMP represents a standard Load Balancer.



For the first experiment we aim to analyze the Flow Completion Time (FCT) of short flows obtained when applying our solution and comparing it to the baseline ECMP. The FCT is measured as the time from when the first packet leaves the source WAN border router where the eBPF splitter is located until the last packet of the flow arrives at the input of the destination WAN border router.

In this analysis, Fig.\ref{fig:performance}a examines the effect of the eBPF splitter's threshold on the FCT acceleration relative to the baseline. The data illustrates that a lower threshold correlates with a significant FCT acceleration, achieving over double the speed for network rates exceeding $\lambda = 34\times10^3$ packets per second. The critical range is between 1 to 40 packets, identified as the typical size for Remote Procedure Calls (RPCs)\cite{aequitas}. Our findings indicate that our method facilitates a speedup even under conditions of minimal network use, and in general managing to boost performance from 1.3 to 2 times compared to employing a load balancer (ECMP).

Fig.~\ref{fig:performance}b, focuses on the impact of network utilization on a determined flow length. It illustrates that as network utilization escalates, longer flows tend to clog the queues with their packet bursts, inducing delays for shorter flows. Nevertheless, through the separation of short and long flows, the former can avoid these extensive packet bursts in queues. This segregation leads to enhanced efficiency as network usage grows, with short flows completing from 1.5 to 2 times faster when using our solution in a 40\% network utilization environment. 


The detailed analysis presented in Fig.~\ref{fig:pdf} focuses on the Probability Density Functions (PDF) analysis of two key metrics: packet jitter and FCT. Regarding packet jitter, the initial two rows of illustrations indicate that our approach enhances both the mean and the standard deviation of jitter. This improvement occurs because our method helps packets avoid queues formed due to long-flow traffic, particularly in congested networks. An examination of the FCT PDF reveals similar outcomes: our strategy surpasses the baseline across all examined cases, proving that it lowers not just the average FCT, but also trims down tail latency in the majority of situations.

\begin{figure}[htbp]
        \includegraphics[width=\columnwidth]{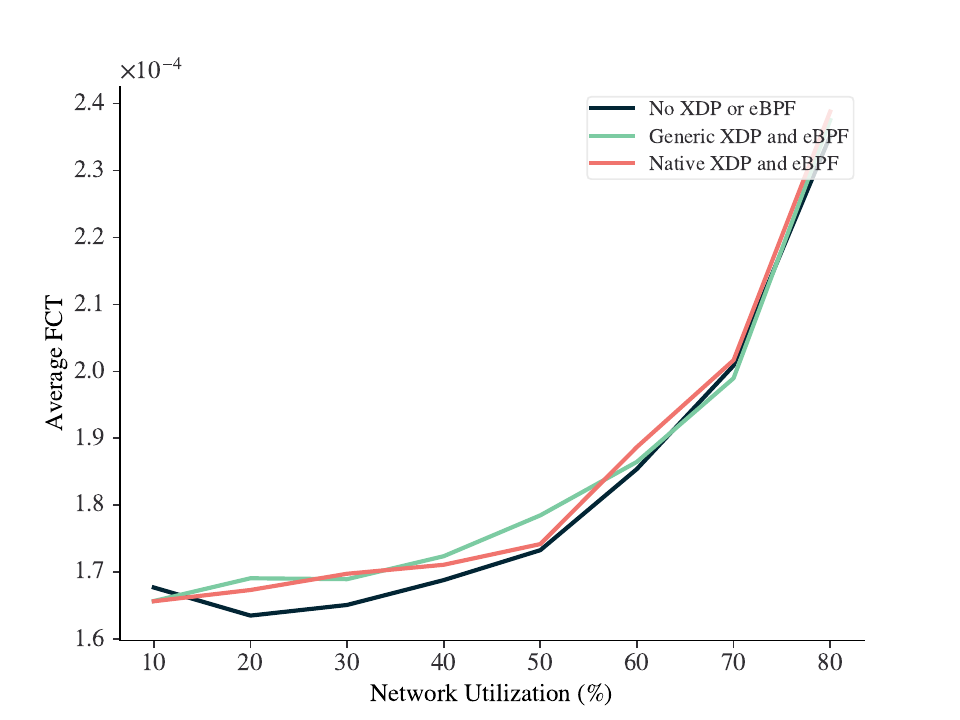}
        \caption{Overhead of eBPF on Flow Completion Time}
        \label{fig:ebpf}
    \end{figure}

The final experiment is designed to explore the overhead that eBPF has on the delay of the flows. The eBPF splitter can introduce some additional delay caused by the additional processing of packets. To assess this impact we measure the FCT of queries of the same size under three distinct conditions \emph{i.} unmodified packets, \emph{ii.} packets processed with eBPF in the kernel as part of the standard network path (Generic XDP and eBPF), and \emph{iii.} packets processed with the eBPF program loaded by the Network Interface Card (NIC) into its initial receive path (Native XDP and eBPF). The results, shown in Fig.~\ref{fig:ebpf}, indicate that the delay introduced is in the order of microseconds, two orders of magnitude lesser than the FCT. If needed, this could be improved even further by using eBPF fully offloaded at the NIC, executing entirely off of the CPU.

\section{Related Work}

Several notable examples have utilized flow length distinction to reduce FCT. For instance, RPCValet \cite{rpcvalet} implements dynamic load balancing across CPU cores based on on-chip network interfaces and real-time load monitoring. PFabric \cite{pfabric} focuses on decoupling flow scheduling from rate control in datacenter transport, also aiming to reduce tail latency. Another recent work, Aequitas \cite{aequitas}, addresses the popularity of distributed microservice architectures by using sender-driven admission control to reduce the FCT of RPCs in datacenter networks.

While these efforts have primarily focused on datacenter networks, where short flows constitute the majority of traffic \cite{aequitas}, the traffic composition changes when merging with WAN traffic, which is mostly composed of long flows. Although Annulus \cite{annulus} recognizes the challenges brought by this mixture of traffic, their solution is focused on developing congestion control mechanisms for datacenter traffic.

In contrast, our work draws inspiration from these previous efforts and aims to apply the lessons learned from datacenter networks to the WAN, especially as applications increasingly move towards hybrid scenarios.

\section{Summary and conclusions}

This paper presents a novel approach to traffic optimization for distributed applications across Wide-Area Networks, drawing inspiration from optimization strategies traditionally applied within datacenters. We focus on addressing the widely recognized performance bottlenecks in the WAN by separating short and long flows and routing them via different tunnels to mitigate congestion. 

Through extensive experimentation, we demonstrate the effectiveness of our proposed solution. Our results consistently show that segregating short and long flows results in substantial improvements in the FCT for short flows. This improvement is achieved without significantly impacting the FCT of long flows, thus ensuring overall application performance remains high.

As applications become increasingly distributed, the role of the network, particularly the WAN, in ensuring optimal application performance is becoming paramount. Our solution addresses the challenges that arise in this context, offering an effective, application-agnostic strategy to optimize network traffic across datacenters.

\section*{Acknowledgments}

This publication is part of the Spanish I+D+i project TRAINER-A (ref.PID2020-118011GB-C21), funded by MCIN/ AEI/10.13039/501100011033. This work is also partially funded by the Catalan Institution for Research and Advanced Studies (ICREA) and and the Secretariat for Universities and Research of the Ministry of Business and Knowledge of the Government of Catalonia aarxivarnd the European Social Fund.

\bibliographystyle{IEEEtran}
\bibliography{paper.bib}

\end{document}